\documentclass[useAMS,usenatbib,a4paper,referee]{mn2e}
\usepackage{graphicx}
\usepackage{amsmath}
\usepackage{txfonts}
\usepackage{mathrsfs}

\def\aap{A\&A}
\def\apjl{ApJ}

\def\apjs{ApJS}

\newcommand{\be}{\begin{equation}}
\newcommand{\ee}{\end{equation}}
\newcommand{\bea}{\begin{eqnarray}}
\newcommand{\eea}{\end{eqnarray}}

\title[Cosmic Chronometers]{Analyzing $H(z)$ Data using Two-point Diagnostics}
\author[Leaf \& Melia]{Kyle Leaf$^{1}$\thanks{kyleaf@email.arizona.edu} and
Fulvio Melia$^{2}$\thanks{John Woodruff Simpson Fellow. E-mail: fmelia@email.arizona.edu} \\
$^1$Department of Physics, The University of Arizona, AZ 85721, USA \\
$^2$Department of Physics, The Applied Math Program, and Department of Astronomy, 
The University of Arizona, AZ 85721, USA}

\begin{document}

\date{}

\pagerange{\pageref{firstpage}--\pageref{lastpage}} \pubyear{2016}

\maketitle

\label{firstpage}

\begin{abstract}
Measurements of the Hubble constant $H(z)$ are increasingly being used
to test the expansion rate predicted by various cosmological models.
But the recent application of 2-point diagnostics, such as $Om(z_i,z_j)$ and 
$Omh^2(z_i,z_j)$, has produced considerable tension between $\Lambda$CDM's 
predictions and several observations, with other models faring even worse. 
Part of this problem is attributable to the continued mixing of truly 
model-independent measurements using the cosmic-chronomter approach, 
and model-dependent data extracted from BAOs. In this paper, we 
advance the use of 2-point diagnostics beyond their current status, and
introduce new variations, which we call ${\Delta}h(z_i,z_j)$, that are more useful for 
model comparisons. But we restrict our analysis exclusively to cosmic-chronometer data, 
which are truly model independent. Even for these measurements, however,
we confirm the conclusions drawn by earlier workers that the data have 
strongly non-Gaussian uncertainties, requiring the use of both ``median"
and ``mean" statistical approaches. Our results reveal that previous
analyses using 2-point diagnostics greatly underestimated the errors,
thereby misinterpreting the level of tension between theoretical predictions
and $H(z)$ data. Instead, we demonstrate that as of today, only Einstein-de 
Sitter is ruled out by the 2-point diagnostics at a level of significance
exceeding $\sim 3\sigma$. The $R_{\rm h}=ct$ universe is slightly favoured
over the remaining models, including $\Lambda$CDM and Chevalier-Polarski-Linder,
though all of them (other than Einstein-de Sitter) are consistent to
within $1\sigma$ with the measured mean of the ${\Delta}h(z_i,z_j)$ diagnostics.
\end{abstract}

\begin{keywords}
{cosmology: large-scale structure of the universe, cosmology: observations, 
cosmology: theory, galaxies}
\end{keywords}

\newpage
\section{Introduction}
$\Lambda$CDM has done reasonably well accounting for a broad range of data 
and is therefore correctly viewed as the current standard model of cosmology 
(see, e.g., Planck Collaboration 2014). But recent analyses of the 2-point 
correlation function of the cosmic microwave background (CMB; Copi et al. 
2015; Melia 2014), as well as the $Om(z_i,z_j)$ and $Omh^2(z_i,z_j)$ diagnostics 
applied to measurements of the Hubble expansion rate $H(z)$ (Shafieloo et al. 
2012; Sahni et al. 2014), appear to have revealed significant tension between its 
predictions and recent measurements (see, e.g., Zheng et al. 2016). 

In this paper, we directly address the problems highlighted by the various
analyses carried out with the $H(z)$ data, which apparently do not confirm
the anticipated transition from early deceleration to more recent acceleration
in the cosmic expansion rate (Jimenez \& Loeb 2002; Moresco et al. 2016a). With 
this type of work, one typically compiles a unified sample of $H(z)$ versus 
redshift measurements based on various approaches, including the determination 
of differential ages using cosmic chronometers (Moresco et al. 2016b) and the 
inference of a characteristic distance scale revealed in baryon acoustic 
oscillations (BAO; see, e.g., Blake et al. 2012). 

However, it is inadvisable to make combined use of BAO measurements with other
data. The BAO approach results in certain sytematic effects that are difficult 
to quantify, certainly without the pre-assumption of a particular model. A BAO 
determination of $H(z)$ relies on both determining the actual baryon acoustic 
peak of a cluster, and the degree of contamination from the Alcock-Pacz\'ynski 
Effect (see, e.g., Melia \& L\'opez-Corredoira 2017). 

To disentangle these effects, one must choose a cosmological model in order to 
determine the overall co-moving acoustic scale, rendering the determination of 
H(z) using BAO data unique to each model. BAO therefore provide
useful data for analyzing a specific model, but a comparative analysis of multiple 
cosmologies requires a careful recalibration of the acoustic scale for each individual
case. In contrast, the differential age method is fundamentally much simpler. It 
has some systematic error based on using the Extended Press-Schechter 
(EPS) approximation (Moresco et al. 2016a), but is heavily based on directly 
observable galaxy properties. Zheng et al. (2016) demonstrated via the
analysis of several cosmological models that the merger of differential-age 
and BAO data resulted in statistically significant discrepencies when compared
to the use of either data set on its own.

In this paper, we therefore restrict our attention to the sample of thirty
model-independent cosmic chronometer measurements of $H(z)$ (Jimenez et al. 
2003; Simon et al. 2005; Stern et al. 2010; Moresco et al. 2012; Zhang et 
al. 2014; Moresco et al. 2015; Moresco et al. 2016a) to carry out a 
comparative test of several cosmological models, using variants of the 
previously introduced two-point correlation functions that may be applied 
to any cosmology, not solely those framed in the context of $\Lambda$CDM. 
The previously defined $Om(z_i,z_j)$ and $Omh^2(z_i,z_j)$ diagnostics are based 
on the presence of an $\Omega_{\rm m} $ parameter in the formulation of the
expansion rate $H(z)$, which restricts their use primarly to $\Lambda$CDM and 
its variations (Sahni et al. 2014). Throughout this paper, we define 
$\Omega_{\rm m}\equiv \rho_{\rm m}/\rho_c$ to be the matter energy density 
$\rho_{\rm m}$ today as a fraction of the critical density $\rho_c\equiv 3c^2H_0^2/8\pi G$.

We will be comparing the observations with the predicted expansion rate in five models, 
including $\Lambda$CDM, $w$CDM, $R_{\rm h}=ct$ (Melia 2007; Melia \& Abdelqader 2009; Melia \& 
Shevchuk 2012; Melia 2016a, 2017), Chevalier-Polarski-Linder (Chevalier \& Polarski 2001;
Linder 2003) and Einstein-de Sitter. Each of these cosmologies has a distinct relationship 
between $H(z)$ and $z$, and some have a parametrization that allows a best-fit model to 
be identified via the optimization of certain free parameters. In previous studies,
such a parametric fitting of the Hubble constant has been used with the $H(z)$ data
to identify the ideal model parameters in $\Lambda$CDM and, in a few cases, a direct
comparison has also been made between $\Lambda$CDM and $R_{\rm h}=ct$, though the
latter has no free parameters for this purpose, so one must necessarily rely on
information criteria to calculate relative likelihoods (Melia \& Maier 2013;
Melia \& McClintock 2015; Wei et al. 2017).

The 2-point diagnostics differ from this parametric fitting approach in several
distinct ways, extending the model comparison in new and statistically meaningful
diretions. Previously, Ding et al. (2015) and Zheng et al. (2016) employed two 2-point diagnostics 
(i.e., $Om[z_i,z_j]$ and ${Om}h^2[z_i,z_j]$), involving $\Omega_{\rm m}$ and $\Omega_{\rm m}^2$, 
applicable to several cosmological models. In these analyses, the 2-point 
diagnostic was calculated based on the $\Omega_{\rm m}$ parameter in $\Lambda$CDM, 
such that it should be consistent with zero if $\Lambda$CDM were the correct model. 
Then, the expectation value of that diagnostic was determined for all the models 
being considered, and the one most consistent with it was judged to be favored 
over the others. Their results, however, were rather inconclusive. Zheng et al. (2016) found that,
while $\Lambda$CDM was the model most favoured by the data among the tested sample, 
it was still nonetheless significantly inconsistent with the measurements, especially on
the basis of ``median statistics" (see, e.g., Gott et al. 2001). Other dark-matter 
parameterizations, such as $w$CDM and CPL (Chevalier \& Polarski 2001; Linder
2003) performed even worse.  

But one must be aware of the fact that in their 
diagnostic, a value of $H_0$ must be chosen a priori, even though there is still
no consensus on what its true value is, given that high-redshift measurements
(e.g., Planck Collaboration 2014) are in tension with $H_0$ measured locally
(e.g., using Type Ia SNe; see Wei et al 2017, and references cited therein). 
In addition, we shall demonstrate in this paper that the statistics of 2-point 
diagnostics has been used improperly in earlier applications, including those
of Ding et al. (2015) and Zheng et al. (2016). Specifically, we shall show that previous applications
of 2-point diagnostics have over-estimated the confidence level of their results,
thereby over-stating the tension between, e.g., the measured $H(z)$ values and 
predictions by $\Lambda$CDM. We will introduce new 2-point 
diagnostics similar in spirit to the previously defined $Om(z_i,z_j)$ and 
$Omh^2(z_i,z_j)$ which, however, are appropriate for a broader range of models, 
including $R_{\rm h}=ct$, that was not included in Zheng et al.'s (2016) study. 
In spite of these improvements to the analysis of 2-point diagnostics,
however, we will confirm the results first demonstrated by Ding et al. (2015)
and Zheng et al. (2016), that the errors reported in conjunction with the
cosmic chronometer data are at least partially non-Gaussian.

Our goals in this paper are threefold. First, we extend the previous work
with 2-point diagnostics to a more comprehensive comparison of various
models. As noted, our approach allows the inclusion of cosmologies whose
parametrization is not based on $\Omega_{\rm m}$. Second, we retrict our 
analysis to the 30 cosmic chronometer measurements to ensure that the 
data are as free as possible of model biases. Third, we present a more
in-depth analysis of the statistics used with 2-point diagnostics and
demonstrate that one must modify the weighted-mean and median statistics 
when using quantitites such as $Om(z_i,z_j)$ and $Omh^2(z_i,z_j)$.
In \S~2, we re-introduce these previously defined diagnostics, and then 
define a similar diagnostic that is applicable to a broader range of models. 
We summarize the data used in this paper in \S~3, and then carry out the model 
comparisons using the new 2-point diagnostic in \S~4. We end with our 
conclusions in \S~5.

\section{Two-point Diagnostics} 
Two-point diagnostics are statistical tools that allow one to comparatively analyze 
measurements in a pairwise fashion. Specifically, they allow one to employ $n$ 
measurements of a particular variable to construct $n(n-1)/2$ comparisons between 
each pair of data. With the 30 cosmic chronometer
determinations of $H(z)$ included in this study, we therefore have 435 comparisons. 
With this method, one can test both how well each pair of points fits a given 
cosmological model, and also determine how closely the stated error bars actually 
fit a normal distribution. Previous work by Zheng et al. (2016) concluded that
the stated errors in the cosmic-chronometer data are strongly non-Gaussian,
based on the application of the $Om(z_i,z_j)$ and $Omh^2(z_i,z_j)$ diagnostics. 
The method employed by Zheng et al., however, was based on the definition of a 
single diagnostic (for which the expectation value is zero in $\Lambda$CDM), then 
to compare that single diagnostic with its expectation value for each model being 
tested. In this work, we have devised separate, but completely analagous diagnostics 
for each cosmology. 

The apparent non-Gaussianity of the data motivates the use of a second approach, pioneered by Gott et al.
(2001), based on ``median statistics," which does not rely on error propagation.
In Gott's method, one takes advantage of the fact that for any single measurement taken
from some truly random distribution function, one has a 50$\%$ chance of that measurement being
above the true median of the underlying distribution, with no assumption concerning
its form. Therefore, if N measurements are taken and ranked by their value, the 
probability that the true median lies between measurements $i$ and $i+1$ is given 
by the binomial distribution, such that 
\begin{equation}
P_i=\dfrac{2^{-N}N!}{i!(N-i)!}\;.
\end{equation} 
Therefore, one finds the number of measurements away from the median one needs to take in order to 
find the 68$\%$ confidence region of the median. We shall demonstrate, however, that this approach 
is not strictly valid for 2-point diagnostics, prompting the use of Monte-Carlo simulations 
with mock data in order to determine the correct number of steps one ought to take away from the 
median to compute the actual confidence range. As detailed below, the expectation value for the 
2-point diagnostic introduced here is zero for the true cosmology. Based on (the modified)
median statistics, the confidence with which each cosmology's median is consistent with 
zero will therefore determine which cosmological model is favoured by the data.

Zheng et al. (2016) found for the $Om(z_i,z_j)$ diagnostic that each cosmology they
tested, i.e., $\Lambda$CDM, wCDM and CPL, had a median inconsistent with zero at more 
than a 68$\%$ confidence level. But we will show that this blanket negative outcome
is due to an improper use of median statistics. Furthermore, while useful, the 
$Om(z_i,z_j)$ diagnostic is suitable only for a cosmology with an $\Omega_{\rm m}$ term. 
Some cosmologies, such as $R_{\rm h}=ct$ and EdS, lack a parameter analagous to 
$\Omega_{\rm m}$, so the $Om(z_i,z_j)$ and $Omh^2(z_i,z_j)$ diagnostics cannot be 
used in general for model selection. The $R_{\rm h}=ct$ model and Einstein-de Sitter
have only one independent parameter, the Hubble constant $H_0$ today. Therefore, we will 
employ a comparable 2-point diagnostic, valid for each cosmology,
defined as 
\begin{equation}
\Delta h(z_i,z_j)\equiv {1\over H_0}\left(\frac{H(z_i)}{E(z_i)}-\frac{H(z_j)}{E(z_j)}\right)\;,
\end{equation}
where the function $E(z)$ is cosmology-dependent. We stress that the actual value of $H_0$ 
does not affect the statistical comparison between the various models, and is used solely to provide 
a uniform scale for the diagnostic $\Delta h(z_i,z_j)$. As such, the measured variance of the Hubble
constant is also irrelevant to the statistical outcome of the $\Delta h(z_i,z_j)$ analysis. So the Hubble
constant may be chosen arbitrarily without affecting the confidence of the final result, and it
is not necessary to use individually optimized values in each case.
For simplicity, we use the value $H_0=70$ km s$^{-1}$ Mpc$^{-1}$ in all the 
figures and reported distributions. In each case, the $\Delta h(z_i,z_j)$ diagnostic 
should be zero if the cosmology being tested is a good match to the data. The models 
we will compare are as follows:

\vskip 0.1in
\begin{table}
 \small
  \caption{Parameter Values Adopted from the Joint Analysis of Betoule et al. (2014), 
including {\it Planck}$+$WP$+$BAO$+$JLA}
  \centering
  \begin{tabular}{lcccc}
&& \\
    \hline
\hline
&& \\
Model & $\Omega_{\rm m}$ & $w_{\rm de}$&$w_0$&$w_a$\\
&&&& \\
\hline
&& \\
$R_{\rm h}=ct$ & --& --& --& -- \\
$\Lambda$CDM & $0.305\pm0.010$& --& --& -- \\
$w$CDM &  $0.303\pm0.012$& $-1.027\pm0.055$& --& -- \\
CPL & $0.304\pm0.012$& --& $-0.957\pm0.124$& $-0.336\pm 0.552$ \\
Einstein de Sitter & 1.0& --& --& -- \\
&& \\
\hline\hline
  \end{tabular}
\end{table}
\vskip 0.1in

\begin{enumerate}
\item The $R_{\rm h}=ct$ universe (a Friedmann-Robertson-Walker cosmology
with zero active mass; Melia 2016a, 2017). This model is based on the 
total equation of state $\rho+3p=0$, where $\rho$ and $p$ are the total
energy density and pressure of the cosmic fluid (Melia 2007; Melia \&
Abdelqader 2009; Melia \& Shevchuk 2012). In this cosmology, there are 
no free parameters once the Hubble constant $H_0$ is used to scale
$\Delta h(z_i,z_j)$: 
\begin{equation}
E^{R_{\rm h}=ct}(z) = (1+z)\;.
\end{equation}

\item Flat $\Lambda$CDM model, with matter and dark-energy
densities fixed by the condition $\Omega_\Lambda=1-\Omega_{\rm m}$
(when radiation is insignificant). In addition to $H_0$, which is
used to scale $\Delta h(z_i,z_j)$, this model has the parameter
$\Omega_{\rm m}$, for which we use the prior value listed in Table 1. 
For this model,
\begin{equation}
E^{\Lambda{\rm CDM}}(z)= \left[\Omega_{\rm m}(1+z)^3+
\Omega_\Lambda\right]^{1/2}\;.
\end{equation}

\item Flat $w$CDM model, with matter and dark-energy densities
fixed by the condition $\Omega_{\rm de}=1-\Omega_{\rm m}$ (again,
when radiation is insignificant). In this case, the dark-energy equation
of state, $w_{\rm de}\equiv p_{\rm de}/\rho_{\rm de}$, in terms of the
dark-energy pressure $p_{\rm de}$, is unconstrained. This model has
two parameters, $\Omega_{\rm m}$ and $w_{\rm de}$, for which we use
the prior values listed in Table~1. Here,
\begin{equation}
E^{w{\rm CDM}}(z)= \left[\Omega_{\rm m}(1+z)^3+
\Omega_{\rm de}(1+z)^{3(1+w_{\rm de})}\right]^{1/2}\;.
\end{equation}

\item The Chevalier-Polarski-Linder model (Chevalier \& Polarski 2001; Linder 
2003), with 
\begin{equation}
E^{\rm CPL}(z)=\left[\Omega_{\rm m}(1+z)^3+(1-\Omega_{\rm m})(1+z)^{3(1+w_o+w_a)}
\exp(-{3w_a z}/[1+z])\right]^{1/2}\;.
\end{equation}
Again, we use the prior values for the parameters indicated in Table~1.

\item Einstein--de Sitter space, which contains only one matter. This model
has no additional parameters once $H_0$ is used to scale $\Delta h(z_i,z_j)$.
In this case,
\begin{equation}
E^{\rm EdS}(z)=(1+z)^{3/2} \;.
\end{equation}

\end{enumerate}

\vskip 0.1in
\begin{table}
 \small
  \caption{Hubble Parameter $H(z)$ measured at 30 different redshifts}
  \centering
  \begin{tabular}{lccl}
&& \\
    \hline
\hline
&& \\
Redshift & $H(z)$ & $\sigma_H$ & Reference\\
&(km s$^{-1}$ Mpc$^{-1}$)&(km s$^{-1}$ Mpc$^{-1}$)& \\
&&& \\
\hline
&& \\
0.07   & 69    &19.6&   Zhang et al. (2014)\\
0.09 & 69 & 12 & Jimenez et al. (2003)\\
0.12& 68.6 & 26.2 & Zhang et al. (2014)\\
0.17 & 83 & 8 & Simon et al. (2005)\\
0.1791 & 75 & 5 & Moresco et al. (2012)\\
0.1993 & 75 & 5 & Moersco et al. (2012)\\
0.2 & 72.9 & 29.6 & Zhang et al. (2014)\\
0.27 & 77 & 14 & Simon et al. (2005)\\
0.28 & 88.8 & 36.6 & Zhang et al. (2014)\\
0.3519 & 83 & 14 & Moresco et al. (2012)\\
0.3802 & 83 & 13.5 & Moresco et al. (2012)\\
0.4 & 95 & 17 & Simon et al. (2005)\\
0.4004 & 77 & 10.2 & Moresco et al. (2016a)\\
0.4247 & 87.1 & 11.2 & Moresco et al. (2016a)\\
0.4497 & 92.8 & 12.9 & Moresco et al. (2016a)\\
0.4783 & 80.9 & 9 & Moresco et al. (2016a)\\
0.48 & 97 & 62 & Stern et al. (2010)\\
0.5929 & 104 & 13 & Moresco et al. (2012)\\
0.6797 & 92 & 8 & Moresco et al. (2012)\\
0.7812 & 105 & 12 & Moresco et al. (2012)\\
0.8754 & 125 & 17 & Moresco et al. (2012)\\
0.88 & 90 & 40 & Stern et al. (2010)\\
0.9 & 117 & 23 & Simon et al. (2005)\\
1.037 & 154 & 20 & Moresco et al. (2012)\\
1.3 & 168 & 17 & Simon et al. (2005)\\
1.363 & 160 & 33.6 & Moresco et al. (2015)\\
1.43 & 177 & 18 & Simon et al. (2005)\\
1.53 & 140 & 14 & Simon et al. (2005)\\
1.75 & 202 & 40 & Simon et al. (2005)\\
1.965 & 186.5 & 50.4 & Moresco et al. (2015)\\
&& \\
\hline\hline
  \end{tabular}
\end{table}
\vskip 0.1in

\section{Observations}
Parameterized cosmological models, such as those considered in this work, can be 
optimized to fit $H(z)$ measurements, of which there exist several types. In addition 
to optimizing the parameters of the standard model, they have been  used to comparatively
analyze the predictions of $\Lambda$CDM with those of other cosmologies. However,
among these $H(z)$ measurements exist those which are model dependent. If a measurement of
$H(z)$ is model dependent, it cannot be used to adequately compare multiple models. As of
today, the cosmic chronometer approach is the only model-independent method of determining 
$H(z)$, and is therefore the method used exclusively in this work. 

\begin{figure}
\centering
\includegraphics[scale=0.5]{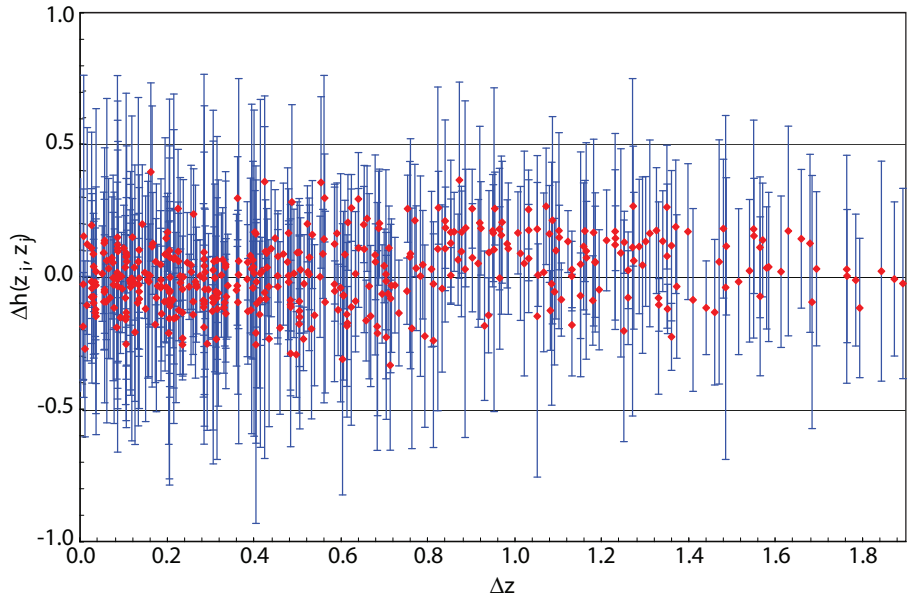}
\label{Fig 1}\caption{The $\Delta h(z_i,z_j)$ 2-point diagnostic calculated
for the sample of 30 cosmic-chronometer measurements of $H(z)$ in the $R_{\rm h}=ct$
cosmology. Red dots denote the calculated central values and the blue bars correspond
to the uncertainties. If the cosmological model is correct, $\Delta h(z_i,z_j)$ should
be zero for all values of $\Delta z$. The equivalent figures for the other cosmologies tested
appear very similar, and are therefore omitted.}
\end{figure}

The cosmic chronometer approach used to determine the data used here constitutes a
differential age method. By comparing the redshift-time derivative ($dz/dt$) of galaxies 
undergoing passive evolution (that is, evolution on timescales much longer than the age 
difference between the galaxies), one can infer the expansion rate H(z) as a function
of redshift. The galaxies used are elliptical and highly massive ($\gtrsim 10^{11}M_{\odot}$). 
They formed most ($\sim 90\%$) of their stellar mass in the redshift range $2<z<3$, over a 
short timeframe of $\sim0.1-0.3$ Gyr. At any given redshift, objects in this morphological 
class are thus the oldest objects in the Universe (Treu et al. 2005), and due to having 
nearly identical evolution history, the ages of stars found in them can be combined with 
the measured redshifts to determine $H(z)$. Specifically, these measurements are based
on the observation of the 4,000{\AA} break in the measured galaxy spectra. Metal absorption 
causes a discontinuity in the spectral continuum, such that the amplitude of the break 
increases linearly with stellar metallicity and stellar age (Moresco et al. 2016a).
If the metal abundance is known, the age difference $\Delta$$t$ between two nearby 
galaxies can be determined by comparing their 4,000{\AA} amplitudes. With $\Delta t$ 
and $\Delta z$ known, one can then find the expansion rate according to 
\begin{equation}
H(z) = -{1\over (1+z)}\,{dz\over dt}
\approx-{1\over (1+z)}\,{\Delta z\over \Delta t}\;.
\end{equation}
The data used in this study are listed in Table~2. These are the same as those
employed by Zheng et al. (2016),  with the exception that their BAO measurements are here
omitted.

\section{Model Comparisons}
Figure 1 shows the complete 2-point diagnostic for all 435 pairs of data for 
$R_h=ct$. The analogous figures for the other models are very similar to this and are
not shown. Note that $\Delta z=z_i-z_j$ is always positive, so that $z_i>z_j$ in
each evaluation of $\Delta h(z_i,z_j)$. The error bars are calculated using standard 
error propagation, assuming the provided errors are consistent with a normal distribution.
The variance of each $\Delta h(z_i,z_j)$ diagnostic for each model is thus given as
\begin{equation}
\sigma_{{\Delta}h_{ij}}^2=\dfrac{{\sigma_i}^2}{{H_0}^2
[E(z_i)]^2}+\sum_x\left[\dfrac{d}{dx}\left(\dfrac{{H(z_i)}}{{H_0}
[E(z_i)]} \right)\sigma_{x}\right]^2+\dfrac{{\sigma_j}^2}{{H_0}^2[E(z_j)]^2}+\sum_x\left[\dfrac{d}{dx}\left(\dfrac{{H(z_j)}}{{H_0}
[E(z_i)]} \right)\sigma_{x}\right]^2\;,
\end{equation}
where ${E(z_i)}$ is the appropriate function for the chosen cosmology and the summation
over $x$ refers to each fitted parameter, such as $\Omega_m$ and $w$ in $w$cdm.
There are no such terms for $R_{\rm h}=ct$ or EdS.  Each $\sigma_x$ value is reported in Table~1 and
$H_0$ is taken to have the uniform value $70$ km s$^{-1}$ Mpc$^{-1}$ in all
cases, as previously noted. The choice of $H_0$ is just for normalization purposes and does not 
affect the statistical outcome in any way. The unweighted distribution of $\Delta h(z_i,z_j)$ values
is shown in figure~2. A careful inspection reveals some departures from a pure Gaussian shape, 
e.g., with tails in figures~2(b-d) and, as we shall quantify below, a greater 
central peaking than one expects for a normal distribution.

Following our discussion above, we study the 2-point diagnostic 
using both weighted-mean and median statistics. The weighted mean formula for the 
$\Delta h(z_i,z_j)$ diagnostic is just
\begin{equation}
{\Delta}h_{\rm w.m.}=\dfrac{\Sigma_{i=1}^{n-1}{\Sigma}_{j=i+1}^{n}{\Delta}h(z_i,z_j)/
{\sigma}_{{\Delta}h_{ij}}^2}{{\Sigma}_{i=1}^{n-1}\Sigma_{j=i+1}^{n}1/{\sigma}_{{\Delta}h_{ij}}^2}\;,
\end{equation}
and its variance would naively appear to follow as
\begin{equation}
{\sigma}_{{\Delta}h_{\rm w.m.}}^2=\left( {\Sigma_{i=1}^{n-1}{\Sigma}_{j=i+1}^{n}1/
{\sigma}_{{\Delta}h_{ij}}^2}\right)^{-1}\;,
\end{equation}
the approach used by Ding et al. (2015) and Zheng et al. (2016) in their analyses. This determination of the 
error is inaccurate, however, and significantly underestimates the true error in the mean of the 2-point diagnostic. We
therefore propose the following approach which, as we shall see, will by construction alleviate much of the tension 
in the results of the previous works.

To illustrate this point, let us suppose that a measured parameter has a mean ${\mu}$ and 
standard deviation ${\sigma}$. Now pull three random measurements from this distribution,
each with its own reported error $\sigma_i$, at increasing redshifts ${z_1}$, ${z_2}$, 
and ${z_3}$. Then, if we construct the 2-point diagnostics consistent with the method
outlined by Ding et al. (2105) and Zheng et al. (2016), and also followed in this paper, we find three values
of the diagnostic for which the lower redshift is subtracted from a higher one, i.e., 
${z_3-z_2}, {z_3-z_1}, {z_2-z_1}$. If we proceed to calculate the unweighted mean of 
these three diagnostic values, we find the following: $\mu_{\rm 2-point}=(2z_3-2z_1)/3$.
Through normal error propagation, this quantity has a standard deviation 
$\sigma=\sqrt{4{\sigma_3}^2+4{\sigma_1}^2}/3$, which is very different from the 
more naiive standard deviation given in Equation~(11), and in fact does not even 
depend on the middle meaurement at all. This happens because taking 
the mean of a 2-point diagnostic incurs a heavy contribution from the highest and lowest 
redshifts, but far less from the middle redshifts in the sample, as they are sometimes 
added and sometimes subtracted when evaluating the 2-point function. Equation~(11) ignores
this eventuality. Therefore, the true method for determining the standard deviation of  
the 2-point diagnostic must involve a careful application of standard error propagation 
to the weighted mean formula. 

The actual calculation of the error in the mean becomes more complex when using 
weighted data. The weighted mean is calculated as in Equation~(10), though one must pay 
attention to the multiplicative terms that affect each of the original $n$ data points. 
One finds that the weighted mean (Equation ~10) can be rearranged into the form
\begin{equation}
{\Delta}h_{\rm w.m.}=\dfrac{\Sigma_{i=1}^{n}{\alpha_i}H(z_i)}{{\Sigma}_{i=1}^{n-1}
\Sigma_{j=i+1}^{n}1/{\sigma}_{{\Delta}h_{ij}}^2}\;,
\end{equation}
with
\begin{equation}
\alpha_i=\Sigma_{j=1}^{i-1}\dfrac{1}{\sigma_{{\Delta}h_{i,j}}^2}-\Sigma_{k=1+i}^{N}\dfrac{1}
{\sigma_{{\Delta}h_{i,j}}^2}\;,
\end{equation}
in which each $\alpha$ is the sum of every term in the numerator of the weighted mean 
that multiplied $H(z_i)$. Note that as with the constant-dispersion case noted above, 
this tends to weight the measurements of H(z) preferrentially at the high and low redshifts, 
while underrepresenting the intermediate values. The true variance of the mean is therefore:
\begin{equation}
{\Delta}{\sigma^2}_{\rm w.m.}=\dfrac{\Sigma_{i=1}^{n}{\alpha^2_i}
\sigma^2(z_i)}{({\Sigma}_{i=1}^{n-1}\Sigma_{j=i+1}^{n}1/{\sigma}_{{\Delta}h_{ij}}^2)^2}\;,
\end{equation}
which turns out to be always greater than the variance naively expected from Equation~(11).

Unlike the weighted mean, which assumes that all errors are Gaussian, median statistics 
makes no a priori assumption concerning the underlying distribution of a measurement 
compared with its true value. The probability that a given observation is above the 
`true' median (i.e., the median of a very large number of measurements) is simply 
calculated by the binomial distribution given in Equation~(1), where $N$ is the 
total number of measurements, and $n$ is the value's position in the distribution. 
The smallest value has $n=1$, while the largest is $n=N$. However, despite making no 
assumptions about the error distribution, Gott's (2001) method does assume that all 
measurements are completely uncorrelated, which cannot be the case in a 2-point measurement, 
where each of the $N$ data points affects the rest of the $N-1$ 2-point values. As such, 
if the binomial distribution predicts that the true median lies within a certain number
of measurements of the median of the 2-point distribution, the actual $68\%$ confidence 
region is significantly greater.
 
To demonstrate this, we have generated mock data sets consisting of 30 determinations 
of H(z), using the measured $H(z)$ values in Table 2 with their reported $\sigma_H$ standard 
deviations, sampling them assuming a normal distribution. These mock $H(z)$ measurements 
were then used to determine a mock 2-point set of 435 diagnostics for each of the models. 
We produced 1 million such mock data samples, recorded their medians, and then determined
the $68\%$ confidence region of the median. Finally, we checked to see how many steps one 
must take away from the median of each mock data set in order to reach that $68\%$ region. 
Following this approach, we have found that the 2-point diagnostic for all the models has 
a $68\%$ chance of lying between the 192nd and 244th measurement (when ranked by value). 
The 218th measurement is the reported median of each set. This is a much larger range
than that simply estimated using Equation~(1), for the simple reason that the 2-point
diagnostics are correlated. The proper use of median statistics in such cases is therefore
not to calculate the confidence region based on a pure binomial distribution but, rather,
to mitigate the impact of correlations by using a Monte-Carlo approach. In this paper,
we report the error above and below the median of each 2-point diagnostic as
the difference between the 218th and 192nd/244th ranked measurements.

Based on the weighted mean approach, using the corrected error estimation in
Equation~(14), we find that the $\Delta h_{\rm w.m.}$ 2-point diagnostic applied to 
$R_{\rm h}=ct$, $\Lambda$CDM, $w$CDM and CPL is consistent with zero to within $1\sigma$ 
in every case (see Table~3). Einstein-de Sitter, however, is ruled out at better than $\sim 5\sigma$.
Furthermore, the value of the weighted mean for EdS is negative, implying that measurements taken
at greater redshift point to smaller $H_0$ values than observations at low redshifts. The
outcomes listed in Table~1 favour $R_{\rm h}=ct$ slightly over the other cosmologies, though
no strong preference can be inferred from the 2-point statistic used here. These results 
therefore weakly confirm earlier conclusions that $R_{\rm h}=ct$ is strongly preferred over 
$\Lambda$CDM based on the use of information criteria to assess the quality of the fits to the $H(z)$ 
data (Melia \& Maier 2013; Melia \& McClintock 2015; Wei et al. 2017). More importantly,
in contrast to earlier reports (e.g., Zheng et al. 2016), we find a much reduced tension 
between theoretical predictions and the actual measurements. As we have already noted,
this difference is entirely due to our more careful handling of the weighted-mean statistics. 

\begin{figure}
\centering
\includegraphics[scale=0.6]{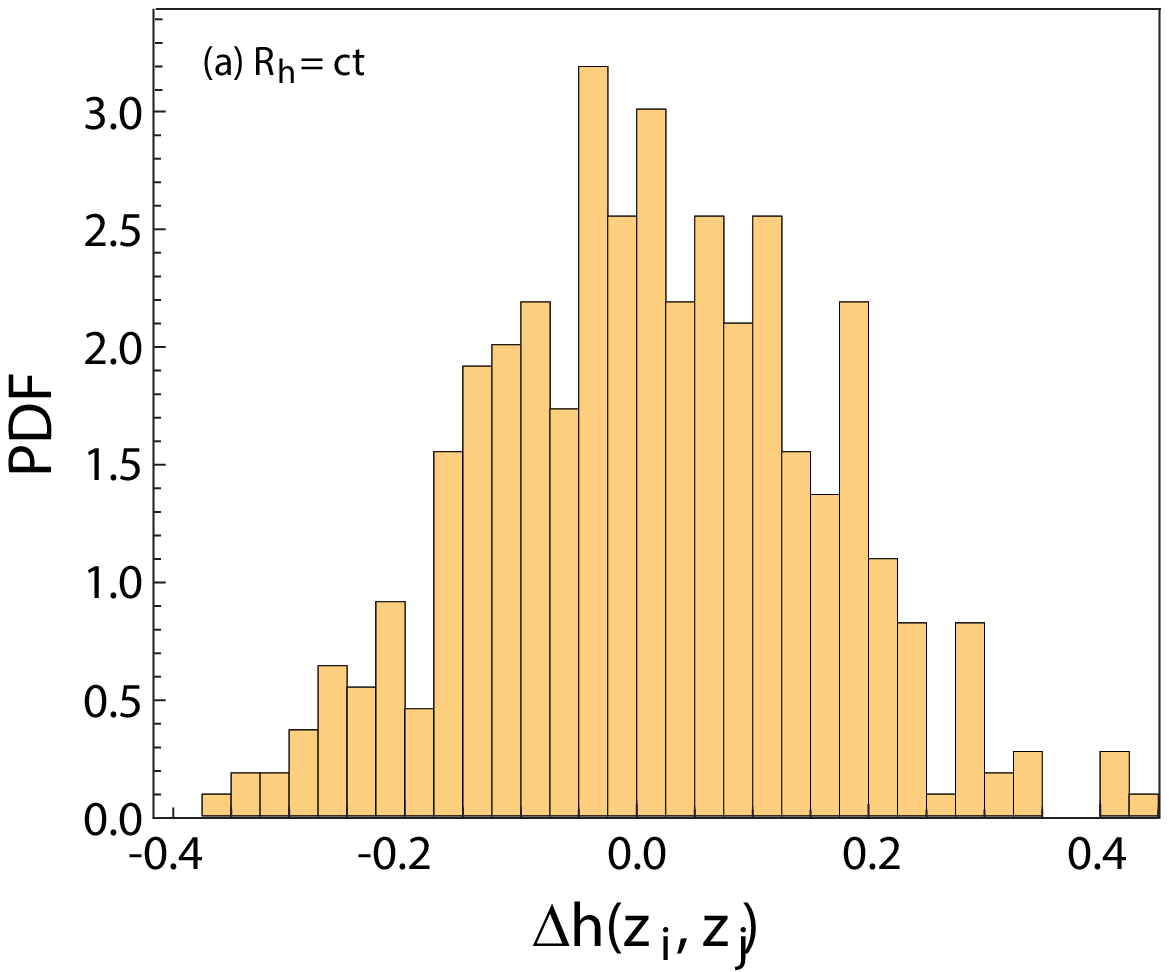}\hskip 0.5in\includegraphics[scale=0.6]{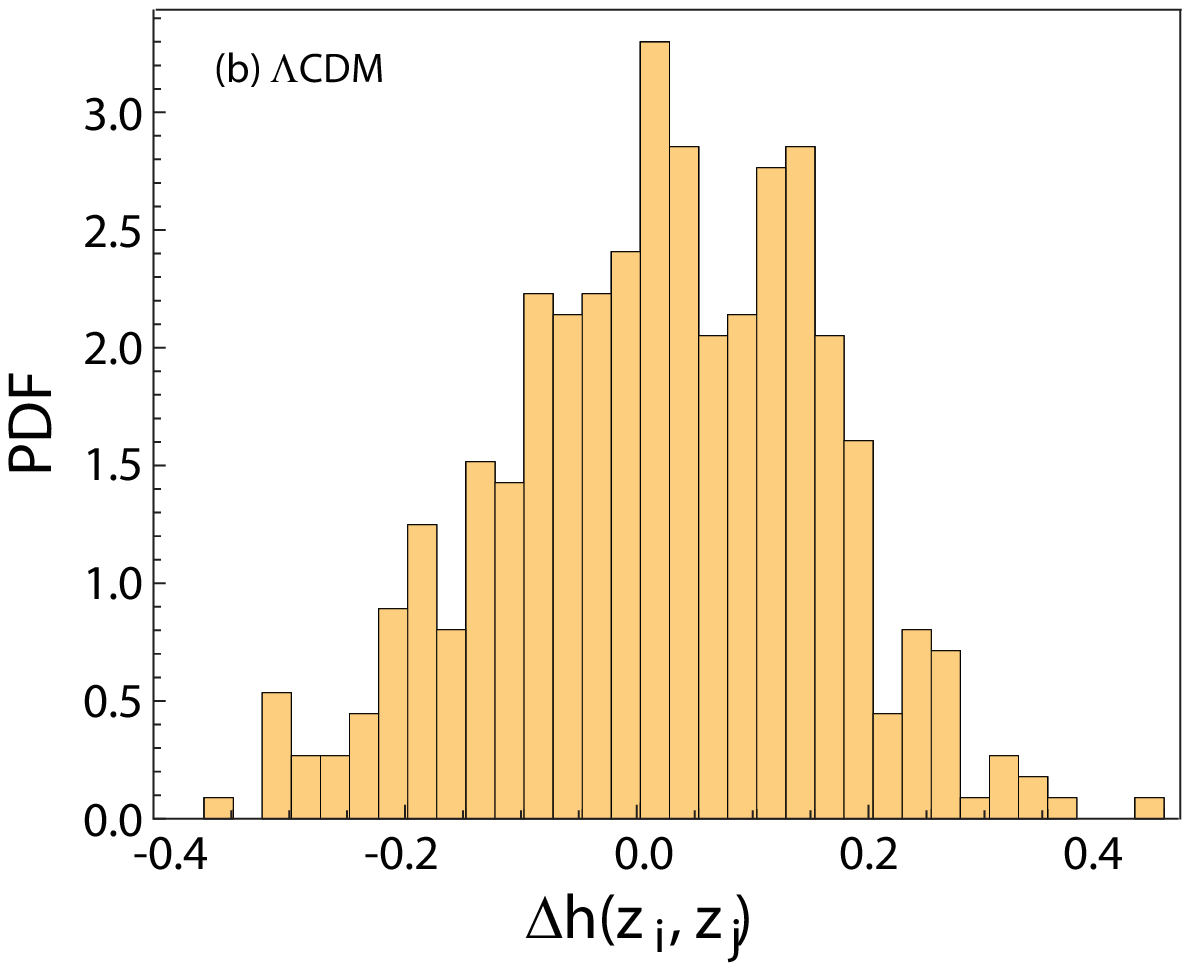} \\
\includegraphics[scale=0.6]{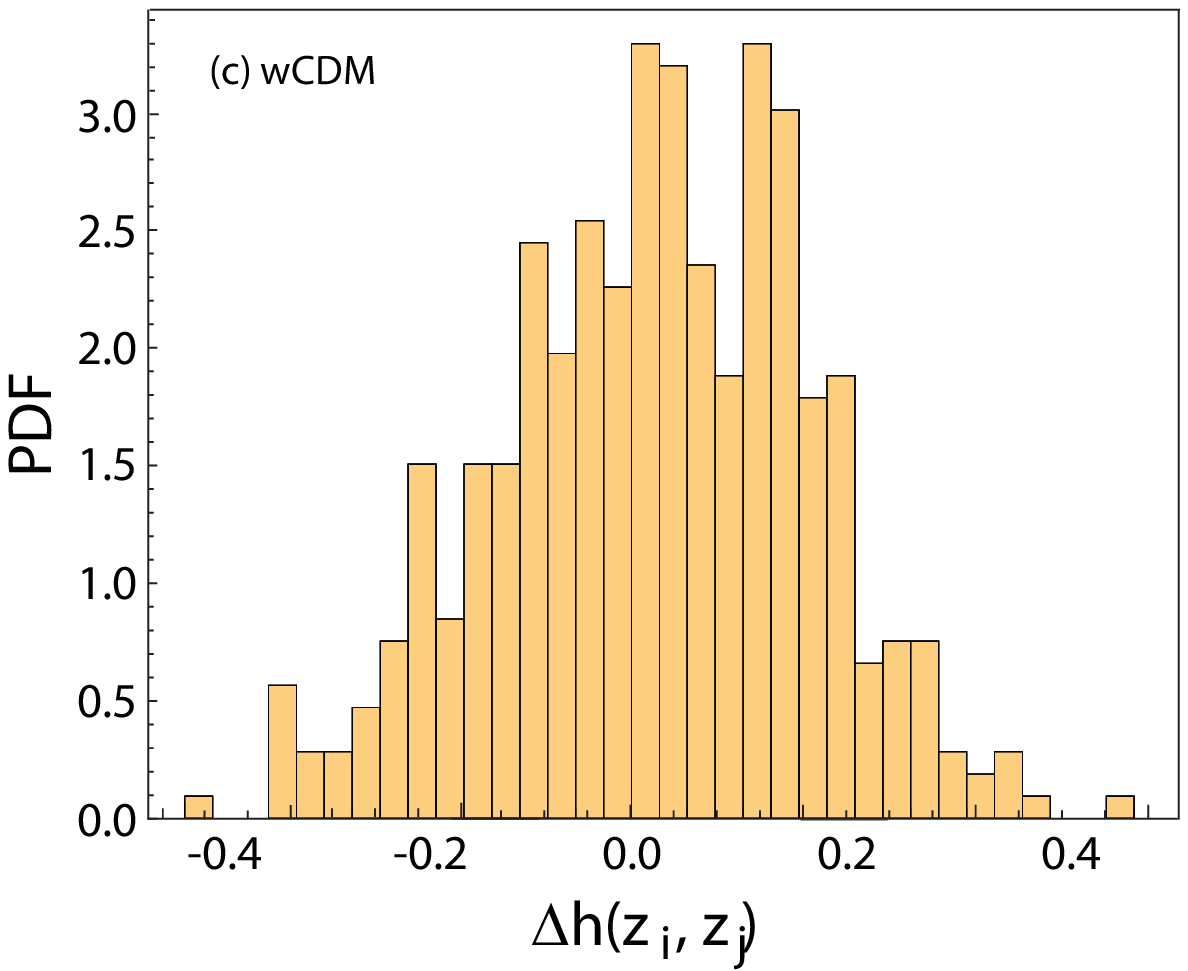}\hskip 0.5in\includegraphics[scale=0.6]{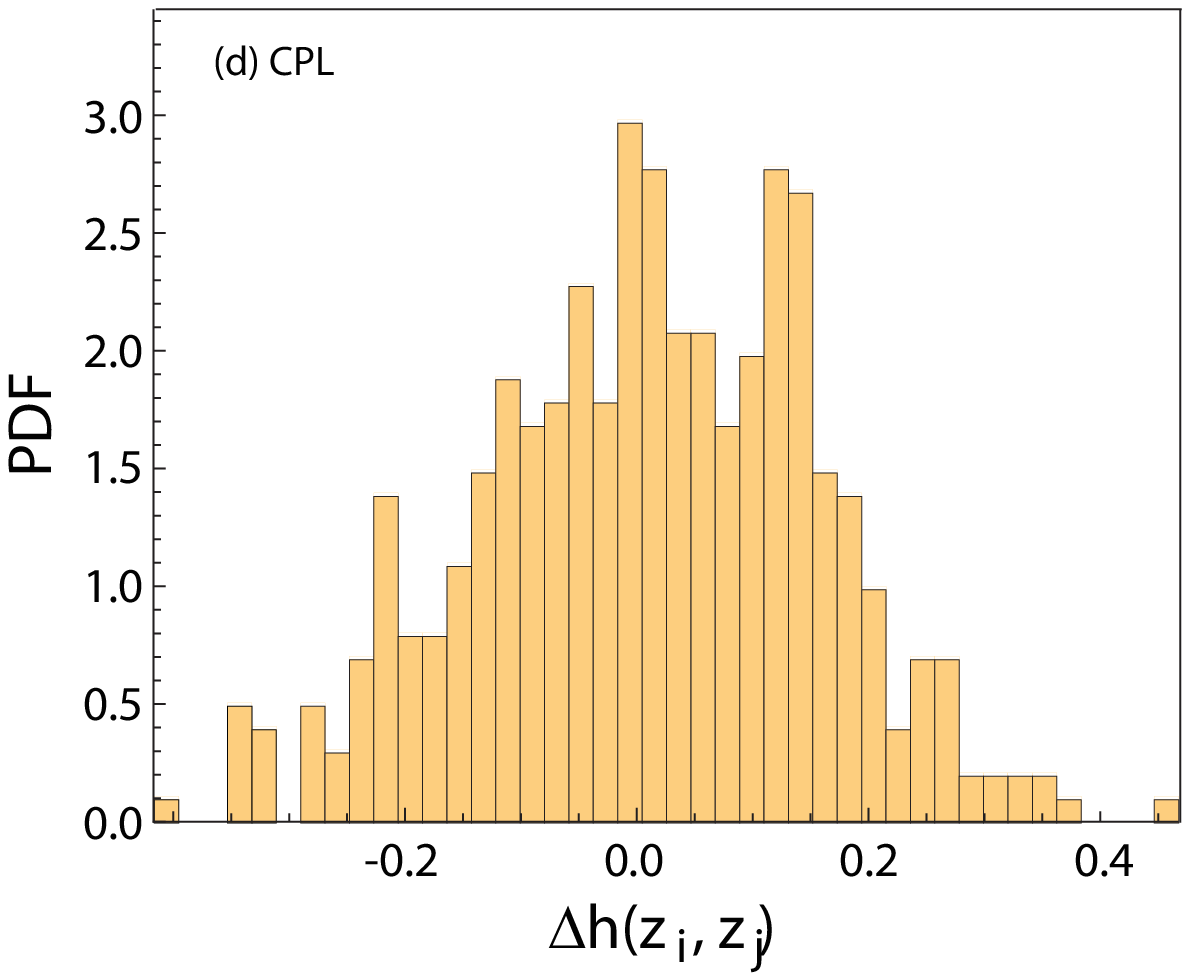} \\
\includegraphics[scale=0.6]{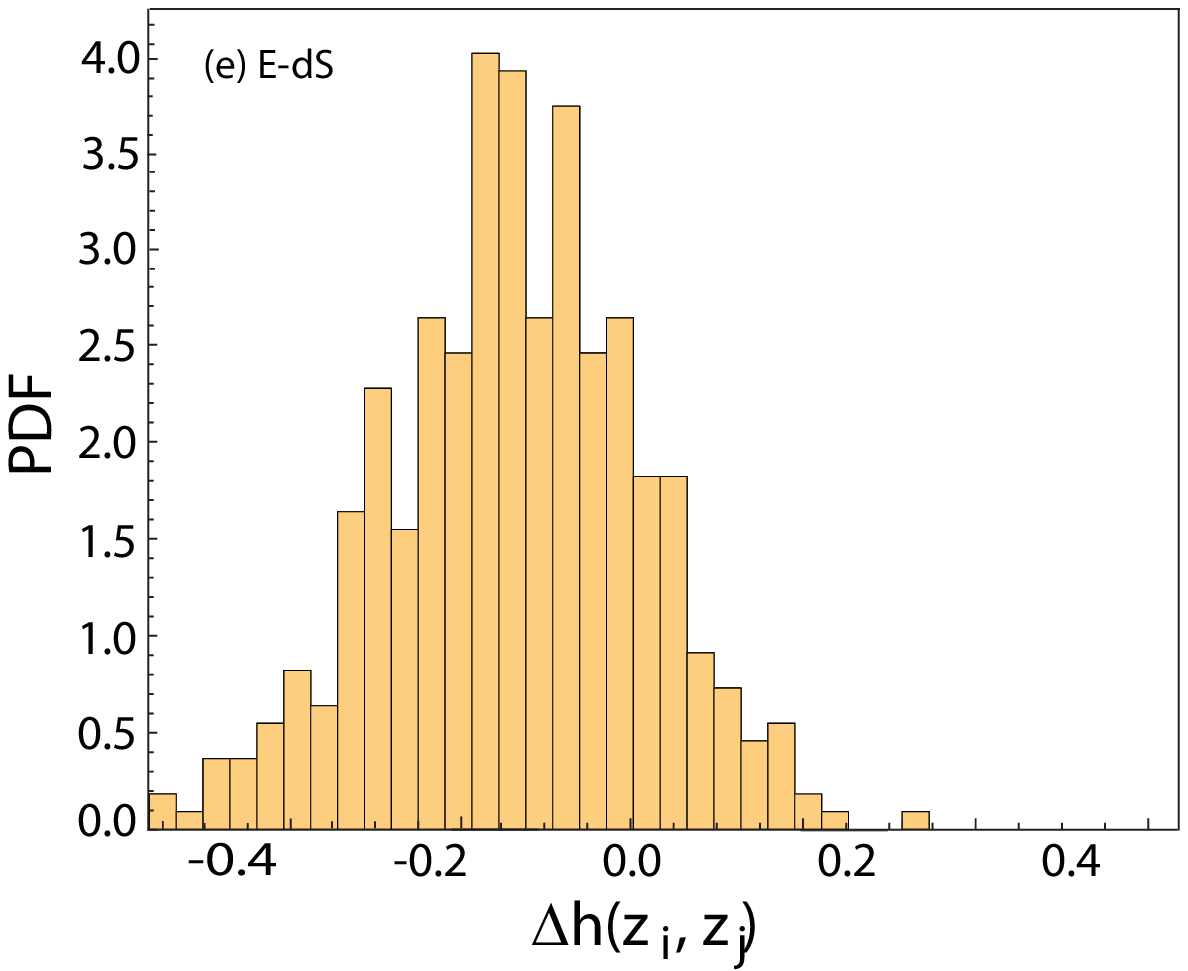}
\label{Fig 2}\caption{Histogram of the $\Delta h(z_i,z_j)$ 2-point diagnostic calculated
for the sample of 30 cosmic-chronometer measurements of $H(z)$, assuming (a) the $R_{\rm h}=ct$
cosmology; (b) $\Lambda$CDM; (c) $w$CDM; (d) CPL; and (e) Einstein-de Sitter.}
\end{figure}

Note also that column 4 in Table~3 indicates a deviation from true Gaussian errors, 
confirming the pioneering analysis by Zheng et al. (2016) in this regard, who used the 
$Om(z_i,z_j)$ diagnostic. 
For all the models we examine here, the percentage of pairs $N_\sigma$ with a 
$\Delta h(z_i,z_j)$ within $1\sigma$ of the weighted mean is greater than the 
expected $68\%$ if the errors associated with the measured $H(z)$ values 
(Table~2) are truly Gaussian. The fact that more measurements of $\Delta h(z_i,z_j)$ lie within 
$1\sigma$ than predicted by Gaussianity (see also fig.~2) suggests that the published 
error bars are generally larger than their actual values, or that there exists some 
generic systematic error that results in most measurements being offset from their 
true value in a similar manner. 

One may already infer this for about half of the 
measurements in Table~2, for which both the statistical $\sigma_{\rm stat}$ and 
systematic $\sigma_{\rm sys}$ contributions to $\sigma_i$ have been reported 
(Gaztanaga et al. 2009; Moresco et al. 2012; Moresco et al. 2016a). In some cases, 
$\sigma_{\rm sys}\sim \sigma_{\rm stat}$, so even if only a portion of $\sigma_{\rm sys}$
is correlated, the overall $\sigma_i$ calculated in quadrature overestimates the
true Gaussian error. Unfortunately, this breakdown in statistical versus systematic
errors is only known for a portion of the sample in Table~2. And even for those
measurements where both $\sigma_{\rm stat}$ and $\sigma_{\rm sys}$ are available,
one does not know which fraction of the latter is correlated. For the study reported
in this paper, we have therefore settled on the simplest approach, which is to
use all of the data in Table~2, with the caveat that the published errors are almost
certainly overestimated. At least this approach ensures consistency across the
entire sample, though the results do confirm that the errors are not perfectly Gaussian.

\vskip 0.1in
\begin{table}
 \small
  \caption{Statistical Outcomes}
  \centering
  \begin{tabular}{lrccrccc}
&& \\
    \hline
\hline
&& \\
Model & $\Delta h_{\rm w.m.}\qquad$ & Offset from &$|N_\sigma|<1$ & $\Delta h_{\rm m.s.}\quad\;$ & $68\%$ Range\\
&&zero (w.m.)&&&(\rm m.s.)\\
\hline
&& \\
 $R_{\rm h}=ct$ & $0.00747{\pm}0.02786$ &$0.27\sigma$&$87.36\% $& $0.00879\quad$&$(-0.01237,+0.02194)$\\
$\Lambda$CDM& $0.01209{\pm}0.03082$ & $0.39\sigma$& $90.11\%$ & $0.02126\quad$& $ (-0.01469,+0.02183)$\\
wCDM& $0.01360{\pm}0.03122$ & $0.4355\sigma$&$90.34\%$ & $0.02528\quad$&$ (-0.01780,+0.02118)$\\
CPL & $ 0.01341{\pm}0.03942$ &$0.3402\sigma$&$ 97.47\%$ & $0.02654\quad$& $(-0.01580,+0.02362)$\\
EdS& $ -0.10851{\pm}0.02146$&$5.06\sigma$&$ 85.06\%$ & $-0.10802\quad$&$(-0.01756,+0.01248)$\\
&& \\
\hline\hline
  \end{tabular}
\end{table}
\vskip 0.1in

Of course, this also implies that the actual error in the weighted mean $\Delta h_{\rm w.m.}$
is probably smaller than what we find here, so the cosmologies we test are likely somewhat 
less consistent with $\Delta h_{\rm w.m.}=0$ than is reported in Table~3. Clearly,
this study will need to be updated once a better handle is available on the nature 
of $\sigma_{\rm sys}$ for $H(z)$ measurements using cosmic chronometers.

Median statistics generally provides a picture consistent with this perspective. 
For Einstein-de Sitter, the result is similar to that of the weighted-mean approach, 
in that the true median in this model is entirely inconsistent with zero. The $R_h=ct$ 
universe is the only model for which a diagnostic value of zero lies within the $68\%$ 
confidence interval. All other models are inconsistent with zero to some degree, although 
$\Lambda$CDM is preferred over $wCDM$ and $CPL$. Notice also the apparent existence of a
double peak in the $\Lambda$CDM, $w$CDM and CPL histograms (figs.~2b, 2c, 2d). As 
noted earlier, the use of median statistics avoids the problem of the weighted mean, in 
not assuming how the data are randomly distributed. As such, results based on
median statistics may be stronger than those derived from the weighted mean. 
Both reject EdS, however, and prefer $R_h=ct$ over other models compared
here. Going forward, we suggest that our method of estimating the error in the 
median using Monte-Carlo to construct mock samples is the proper way of finding
 the error associated with 2-point diagnostics, rather
than the binomial distribution used by Zheng et al. (2016) and others, given
that the latter significantly underestimates the errors. 

\section{Conclusion}
The previously introduced 2-point diagnostics, such as $Om(z_i,z_j)$, have been used
successfully to test the viability of various cosmologies based on the measurement of
$H(z)$. Here, we have extended this work by introducing a new 2-point diagnostic, 
$\Delta h(z_i,z_j)$, which is more generally applicable to a wider range of models,
including those whose formulation does not include the normalized matter density
$\Omega_{\rm m}$. It allows for the analysis of $n(n-1)/2$ pairs of data, offering
different statistics than is available solely with methods that rely on just $n$ 
measurements. 

This approach, however, also introduces some combinatorial effects that 
must be properly accounted for in order to draw accurate conclusions. 
The diagnostic we have introduced here allows for a fair test to determine 
whether the errors associated with the data are being estimated correctly. 
Earlier uses of $H(z)$ measurements included some model-dependent data, namely 
those based on BAO peaks in the galaxy distribution. But most of these measurements 
must pre-assume a cosmology in order to disentangle the BAO peak position from 
redshift-space distortions (RSD) due to internal galaxy motions. As such, BAO 
data tend to be incompatible with all models other than the one used to remove 
the RSD, and are therefore not useful when comparing different cosmologies for 
model selection, unless each model was carefully considered separately from the 
beginning, which has not been the case with existing data. In this paper, we 
have avoided all such biases, relying instead on cosmic chronometer observations, 
which tend to be independent of any model.

Using standard weighted-mean statistics, we have found that the $\Delta{h}(z_i,z_j)$ 
diagnostic is consistent with zero to within $1\sigma$ for all the tested
models, except for Einstein-de Sitter, which is ruled out at over $5\sigma$. The 
caveat with this result is that our analysis has also demonstrated that the errors 
reported for $H(z)$ are not purely Gaussian. The true errors are almost certainly smaller 
than those published, or have a correlated systematic effect that would result in
cosmologies having a diagnostic $\Delta h_{\rm w.m.}$ somewhat less consistent with zero
than what we are reporting here.

Median statistics, based on the use of Monte-Carlo methods to estimate the confidence
region associated with the measured median, rather than simply using a binomial 
distribution that assumes uncorrelated data, has yielded a result consistent with 
that of the weighted mean. Einstein-de Sitter is again ruled out strongly, with 
$R_h=ct$ the significantly preferred model of those tested here. However, the 
confidence intervals of the median of these 2-point diagnostics do not constitute a 
definitive rejection of any model other than EdS.

This result is intruiging, especially when contrasted with the analysis of the same data 
using the $Om(z_i,z_j)$ diagnostic (Ding et al. 2015; Zheng et al. 2016). These 
earlier works used an improper statistical analysis, and did not include the $R_{\rm h}=ct$ 
universe in their comparisons, principally because the $Om(z_i,z_j)$ diagnostic 
cannot be applied to it directly. Zheng et al. (2016) found that, while $\Lambda$CDM 
is favoured over CPL and $w$CDM, even $\Lambda$CDM itself has a 2-point diagnostic 
that is strongly incompatible with zero. On the other hand, we have shown in this 
paper that 2-point diagnostic errors reported in earlier work were severely underestimated.

The remaining issue is whether the results we have derived here carry over
to an analogous statitical analysis using the $Om(z_i,z_j)$ and $Omh^2(z_i,z_j)$ diagnostics.
We have therefore repeated the work of Zheng et al. (2016) using these 2-point diagnostics,
both with their error methodology and our improved treatment that correctly accounts for
correlations in the data. Given that $\Omega_{\rm m}$ is not a model parameter in all
cosmologies, however, we have restricted this comparison to $\Lambda$CDM only. To
summarize the results, we have found that, while the $Om(z_i,z_j)$  diagnostic was 
inconsistent with its expected value of $0$ at  $2.8\sigma$ based on median statistics 
and Zheng et al.'s incorrect error assessment, it is actually inconsistent with this value 
at only $0.96\sigma$ when the errors are estimated correctly using our approach. With
the use of weighted mean statistics, this diagnostic was found by Zheng et al. to be 
inconsistent with $0$ at $0.12\sigma$, whereas the correct error handling yields an
inconsistency at only the $0.04\sigma$ level. The former comparison is more valid
in this case, however, since we all agree that the reported errors are non-Gaussian.
Using median statistics, Zheng et al. also found that the measured $Omh^2(z_i,z_j)$ 
diagnostic is inconsistent with its expected value of $0.1426$ at $4.4\sigma$, while
our corrected error assessment improves this to an inconsistency of $1.5\sigma$.

The statistical analysis of the data using our $\Delta{h}(z_i,z_j)$ diagnostic
therefore appears to by completely consistent with the results based on the use
of $Om(z_i,z_j)$ and $Omh^2(z_i,z_j)$. The advantage of the former, however,
is that it can be used for all cosmologies, not only those in which $\Omega_{\rm m}$
is a free parameter. Very importantly, we have confirmed that our improved
error analysis significantly modifies the conclusions regarding which models are
ruled out by the cosmic chronometer data, irrespective of which 2-point diagnostic
is used in the model comparisons. Based solely on these diagnostics, including
$\Delta{h}(z_i,z_j)$, only Einstein-de Sitter is ruled out strongly by these
observations, though $R_{\rm h}=ct$ is slightly preferred compared to the rest.

\section*{Acknowledgments}
We are grateful to Rabindra Bhattacharya for assistance in finding the proper expression 
for the weighted mean of the 2-point function. We are also grateful to the referee,
Marek Biesiada, for his very thoughtful review, which has resulted in several notable 
improvements to the manuscript.  FM is grateful to the Instituto de Astrof\'isica 
de Canarias in Tenerife and to Purple Mountain Observatory in Nanjing, China
for their hospitality while part of this research was carried out. FM is also 
grateful for partial support to the Chinese Academy of Sciences Visiting 
Professorships for Senior International Scientists under grant 2012T1J0011,
and to the Chinese State Administration of Foreign Experts Affairs under
grant GDJ20120491013.

\label{lastpage}

\end{document}